\documentclass[%
 reprint,
 amsmath,amssymb,
 aps,
floatfix,
]{revtex4-1}

\usepackage{graphicx}
\usepackage{dcolumn}
\usepackage{bm}

\begin{document}

\preprint{APS/123-QED}

\title{Tuning thermoelectric efficiency of polyaniline sheet by strain engineering}

\author{Sahar Izadi Vishkayi}
\author{Meysam Bagheri Tagani}%
 \email{$m{\_}bagheri@guilan.ac.ir$}
\affiliation{%
 Department of Physics, Computational Nanophysics Laboratory (CNL), University of Guilan, Po Box:41335-1914, Rasht, Iran 
}%

\date{\today}

\begin{abstract}
Two-dimensional polyaniline monolayer ($C_3N$) has been recently synthesized as an indirect semiconductor with high electron mobility. In this research, with combination of density functional theory and Green function formalism, we investigate electrical and thermal properties of $C_3N$ sheet in details. It is observed that a tensile strain along zigzag direction can induce a transition from indirect to direct semiconductor, whereas the sheet transits from semiconductor to metal under compressive strain. Thermoelectric efficiency of unstretched $C_3N$ sheet is higher in p-doping and its maximum value is obtained when the transport is along zigzag direction. A reduction in figure of merit is found upon applying strain independent from its direction. To overcome the reduction, we show that when the electrical transport and strain are perpendicular to each other, thermoelectric efficiency of the $C_3N$ sheet can be significantly increased dependent on the  kind of strain (tensile or compression). Results predict the potential application of $C_3N$ sheet in thermoelectric and optoelectronic industry by strain engineering.
\end{abstract}

\maketitle


\section{\label{sec:level1}Introduction}

\par

Discovery of graphene in 2004 \cite{graphene} opens new doors for research in solid state physics and material science. Unique properties of graphene like high electrical and thermal conductivity \cite{graphene, Th}, ballistic transport \cite{BT}, high electron mobility \cite{mob} make it among interesting advanced materials. Integer and half-integer quantum hall effect \cite{QHF, HQHF}, and topological insulator properties observed in the graphene more reveal its high potential. Most important challenge to use the graphene in electronic is absence of the band gap making it unsuitable for field effect transistors. Thereafter, silicene and germanene as counterparts of the graphene were theoretically predicted and experimentally synthesized \cite{si1, ge1}. More single element monolayers has been synthesized in recent years having individual properties. Borophene as a monolayer of boron atoms is a metal with different allotropes \cite{broph}. Antimonene is an indirect semiconductor with ability of tuning its band gap with strain engineering \cite{ant}. Story of two-dimensional (2D) materials enters a new era after discover of transition metal dichalcogenide monolayer (TMD). Unlike graphene, they cover a wide spectrum of band gap and can be used in light emitting diode and optoelectronic industry.
\par In 2016 Mahmood and coworkers prepares a 2D polyaniline sheet via the direct pyrolysis
of hexaaminobenzene trihydrochloride single crystals in solid state \cite{Mahmood}. Synthesized structure has empirical formula of $C_3N$ and a hexagonal lattice composed of six carbon and two nitrogen atoms. Unlike other graphical carbon nitride, the prepared sheet is hole free. Yang et al. \cite{Yang} showed that $C_3N$ sheet is an indirect semiconductor with band gap of $0.39 eV$ and can be changed to a magnetic sheet by partial absorption of hydrogen atoms. Mechanical properties of $C_3N$ sheet has been theoretically studied \cite{Zhou1,Mortazavi,Shi}. In addition, the ability of the sheet for application in lithium-ion battery has attracted some attention \cite{Ion1, Ion2}. We showed that $C_3N$ nanoribbons suffer a transition from semiconductor to half-metal under external electric field and a transition from magnetic metal to semiconductor by passivation of the edge atoms\cite{Bagheri1}. Furthermore, it is shown that a fully hydrogenated $C_3N$ sheet is an insulator with band gap of $5 eV$ \cite{Bagheri2}. Due to similarity with graphene and having band gap, $C_3N$ sheet is a suitable candidate for next-generation electronic devices.
\par Reduce of thermal energy loss in devices is an important challenge in science and industry. Good thermoelectric materials can economize energy waste and increase efficiency. Thermoelectric efficiency of a material is measured by a dimensionless quantity as figure of merit , ZT, given by:

\begin{equation}
ZT=\frac{S^2GT}{\kappa_e+\kappa_{ph}},
\end{equation}

Where $S$ is thermopower, $G$ is electrical conductance and $T$ denotes operating temperature. $\kappa_{ph}(\kappa_{e})$ is lattice (electron) thermal conductance. A lot research has been performed on the thermoelectric properties of nanostructures like quantum dots \cite{QD1, QD2}, quantum wires\cite{Wi1}, and monolayers \cite{Se1, Se2}. Thermal conductivity of $C_3N$ sheet and bilayer has attracted a lot attention in recent years~\cite{new1,new2,new3,new4,new5,new6}. Mortazavi investigated thermal conductivity of $C_3N$ sheet with combination of DFT method and molecular dynamics simulations \cite{new1}. He and coworkers showed that how one can tune themal conductivity of the sheet by adding or removing carbon atoms \cite{new2}. Hong et al. computed thermal conductivity of monolayer and bilayer $C_3N$ sheets using classical molecular dynamics \cite{new3}. Kumar et al showed that thermal conductivity of $C_3N$ sheet is ultralow in comparison to graphene case \cite{new4}.  In this research, we investigate thermoelectric properties of $C_3N$ sheet in linear response regime by combination of density functional theory and Green function formalism. It was shown that thermal conductivity of $C_3N$ sheet is one order lower than graphene case so it is suitable for application in thermoelectric devices. Thermopower, electrical conductance, lattice thermal conductance and figure of merit are analyzed as a function of strain. Strain can be used as an important factor to tune the band gap of the sheet. We show that how tensile and compressive strain can affect thermoelectric efficiency of the sheet. In addition, results reveal that the thermoelectric efficiency is strongly dependent transport and strain direction so that there is a noticeable increase in ZT when strain and transport directions are perpendicular to each other. Our analysis reveals that strain direction, kind of strain (tensile or compressive strain), and transport direction (along zigzag or armchair direction) can significantly modulate thermoelectric performance of $C_3N$ sheet. The article is organized as follows: simulation details and theoretical background are presented in next section. Section 3 is devoted to electrical and thermoelectric properties of $C_3N$ sheet under strain. And some sentences are given as a summary at the end of the article.

\begin{figure}[ht]
\includegraphics[height=80mm,width=85mm,angle=0]{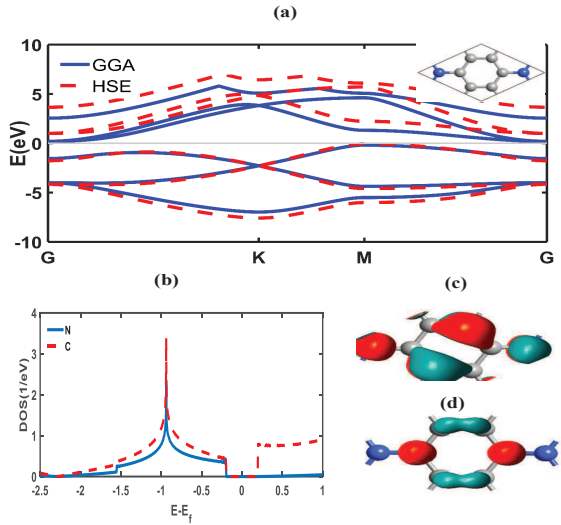}
\caption{\label{band_pure}  (a) Band structure of $C_3N$. GGA based results are shown by solid line and  dashed line is devoted to HSE calculations.
Inset shows a unit cell of $C_3N$ sheet. Carbon and nitrogen atoms are shown by gray and blue balls, respectively. (b) Projected density of states for carbon and nitrogen atoms. (c) and (d) valance and conduction bloch states, respectively. }
\end{figure}

\begin{figure}[ht] \label{gap_strain}
\includegraphics[height=80mm,width=85mm,angle=0]{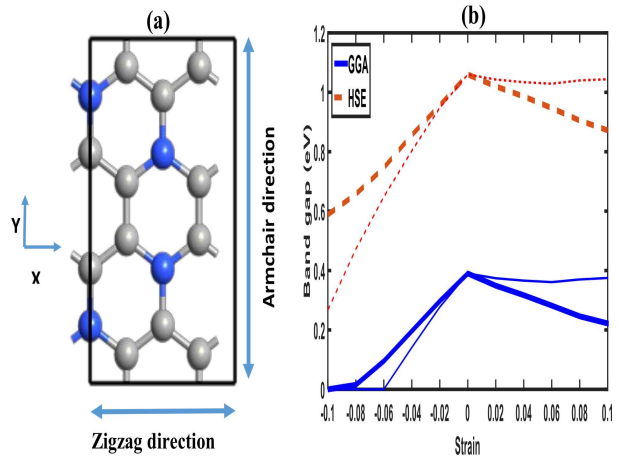}  
\caption{\label{gap_strain}  (a) Conventional unit cell used for strain calculations. (b) Variation of band gap versus strain along zigzag direction (thick lines) and armchair directions (thin lines). }
\end{figure}

\section{\label{sec:level2}Computational method}

\par Electronic calculations are performed using density functional theory implemented in SIESTA package \cite{siesta}. Norm-conserving Troullier-Martins pseudopotential \cite{TM} is used to describe core electrons and Perdew-Burke-Ernzerhof (PBE) generalized gradient approximation (GGA)\cite{PBE} is employed as exchange-correlation functional. Cut-off energy is $100H$ and a $50 \times 50 \times 1$ k-sampling is used to mesh Brillouin zone of a $C_3N$ sheet. Double-zeta single polarized (DZP) basis set is used to describe valance electrons. With respect to the fact that the GGA approximation underestimates the band gap of semiconductors, Heyd-Scuseria-Ernzerhof (HSE) hybrid exchange-correlation functional\cite{HSE} is also used to measure the band gap of samples. All considered structures are optimized so that the force on each atom is less than $0.001 eV/{\AA}$.

\par To investigate mechanical properties of $C_3N$ sheet, a rectangular supercell composed of 16 atoms is considered. For each strain, lattice constant perpendicular to strain direction and position of atoms have been optimized so that force on each atom and stress of the lattice were less than $0.001eV/{\AA}$, and $0.001GPa$, respectively. Brillouin zone of the supercell was meshed by a $40\times 60\times 1$ k-point sampling. Energy dependent electronic transmission of the sheet for each strain is obtained by counting electronic bands crossing each energy. This method is corresponding to obtaining transmission coefficient from nonequilibrium Green function formalism when the structure is perfect and electrodes and scattering region are the same like this study.

\par To compute phonon band structure and phonon transmission coefficient, empirical Tersoff potential\cite{Tersoff} parametrized by Matsunga et al.\cite{MFM}  has been employed. Comparison between phonon band structure obtained from DFT method and classical approach (Fig. S1) shows that the Tersoff potential can produce phonon band structure with suitable quality. For computing phonon transmission coefficient, we have used nonequilibrium green function formalism and defined a central region with length of $4 nm$ coupled to two electrodes. Dynamical matrix is obtained from empirical Tersoff potential and retarded Green function is given as:
\begin{equation}\label{Eq1}
G^{r}_{ph}(\omega)=[\omega^2I-K_c-\Sigma^{r}_{ph,L}(\omega)-\Sigma^{r}_{ph,R}(\omega)]^{-1},
\end{equation}
where $K_c$ is mass-weighted force constant of the central region. $\Sigma^r_{ph,L(R)}$ denotes the self-energy due to coupling between central region and left (right) electrode.

\begin{figure}[hb] \label{band_direction_strain}
\includegraphics[height=90mm,width=85mm,angle=0]{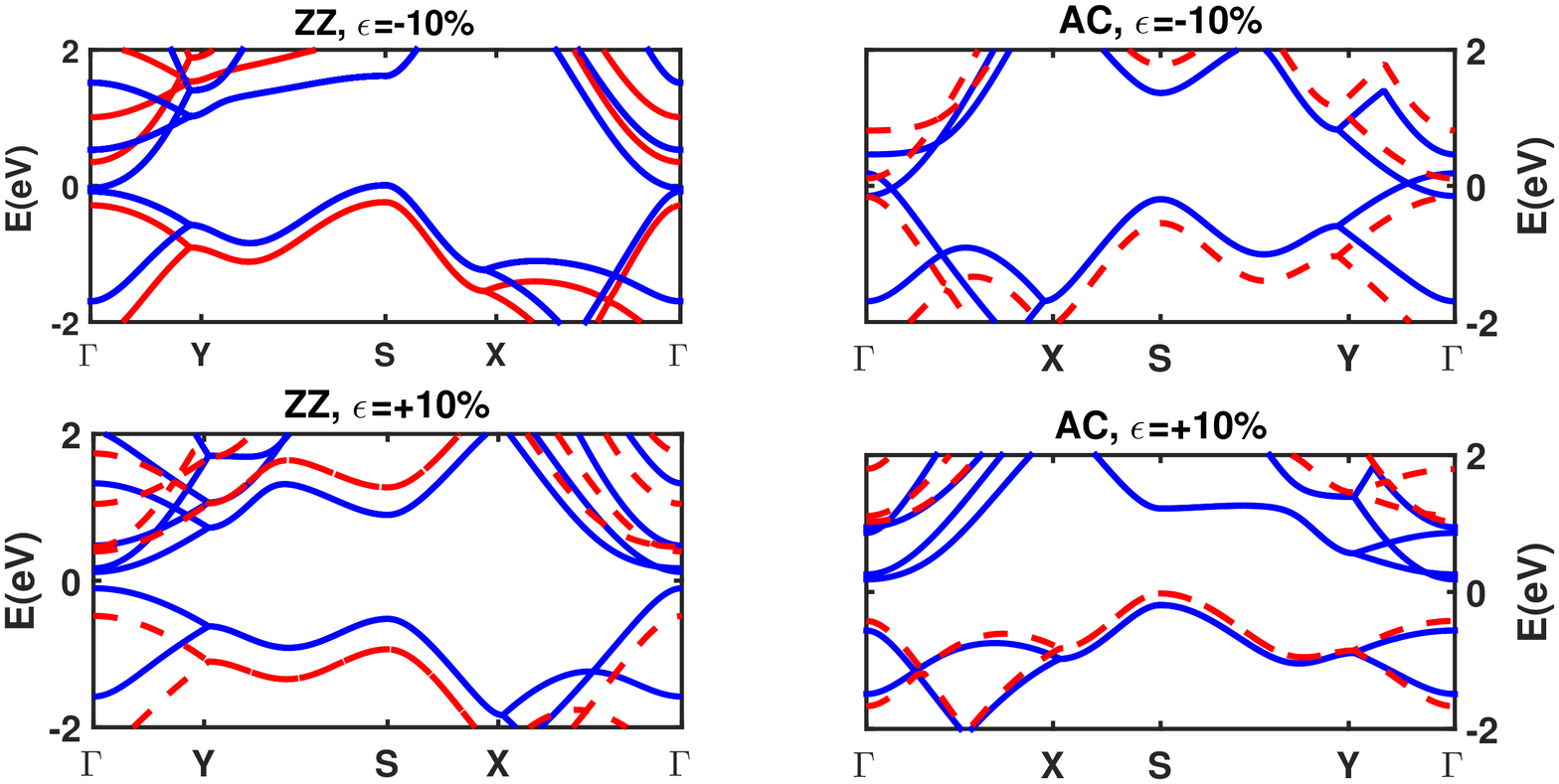}  
\caption{\label{band_direction_strain}  Band structure of $C_3N$ sheet under different strains. GGA and HSE results are shown by solid and dashed lines, respectively. }
\end{figure}

\par Thermoelectric properties of the sheet are investigated in linear response regime and by using of Landau formula. Electrical and thermal currents are given as follows:
\begin{equation}\label{Eq2}
    I_E=\frac{2e}{h}\int{dET_{e}(E)[f_{L}(E)-f_{R}(E)]},
\end{equation}
\begin{equation}\label{Eq3}
    I_Q=\frac{2e}{h}\int{dET_{e}(E)[f_{L}(E)-f_{R}(E)](E-E_f)},
\end{equation}
where $T_e(E)$ is electron transmission coefficient and $f_{L, (R)}$ is Fermi distribution function of left (right) electrode. By expanding the Fermi distribution function in terms of potential and temperature up to first order, one can obtain thermoelectric coefficients in the linear response regime. Electrical conductance, thermopower, and electrical thermal conductance are given as:
\begin{equation}\label{Eq4}
G=e^2L_0,
\end{equation}
\begin{equation}\label{Eq5}
S=\frac{\Delta V}{\Delta T}|_{I=0}=\frac{L_1}{eTL_0},
\end{equation}
\begin{equation}\label{Eq6}
\kappa_e=\frac{1}{T}(L_2-\frac{L_1^2}{L_0}),
\end{equation}
where
\begin{equation}\label{Eq7}
L_n=\frac{2}{h}\int{dE(E-E_f)^nT_{e}(E)(-\frac{\partial{f}}{\partial{E}})}.
\end{equation}
Landau formula for phonons is:
\begin{equation}
I_{ph}=\int_{0}^{\infty}{d\omega \frac{\hbar\omega}{2\pi}T_{ph}(\omega)[n(\omega,T_L)-n_{B}(\omega,T_R)]},
\end{equation}
and phonon thermal conductance in the linear response regime is given as:
\begin{equation}\label{Eq9}
\kappa_{ph}=\int_{0}^{\infty}{d\omega\frac{(\hbar\omega)^2}{2\pi k_BT^2}T_{ph}(\omega)\frac{exp(\frac{\hbar\omega}{k_BT})}{(exp(\frac{\hbar\omega}{k_BT})-1)^2}}.
\end{equation}
Thus, thermoelectric efficiency of the structure is computed as:
\begin{equation}\label{Eq10}
ZT=\frac{S^2GT}{\kappa_{ph}+\kappa_e}
\end{equation}
In following section, we present electrical and thermoelectric properties of $C_3N$ sheet under different conditions.

\section{\label{sec:level3}Results}
\subsection{Electronic properties}
Fig.\ref{band_pure}a shows band structure of a $C_3N$ monolayer based on both GGA and HSE approximations. $C_3N$ sheet is a semiconductor so that valance band maximum (VBM) is located at $M$ high symmetry point and conduction band minimum (CBM) is located in $\Gamma$ point. Energy band gap, $E_g$, is equal to $0.39 eV$ based on GGA approximation. It is well-known that GGA underestimates the band gap. HSE based band structure predicts a higher band gap of $1.06 eV$. Our results are consistent with pervious experimental and theoretical reports \cite{Yang, Zhou1}. The VBM is located near the Fermi level while the CBM is far from it, so the $C_3N$ sheet can be considered as a p-type indirect semiconductor. There is a Dirac point in energy of $-2.3 eV$ in $K$ high symmetry point of Brillouin zone. Position of the Dirac cone is the same as graphene sheet but it is located in negative energies due to existence of nitrogen atoms with five valance electrons in the structure. The Dirac cone we confirmed in experiment of Mahmood et al.\cite{Mahmood}. Unlike the electronic band gap, the position and slope of the Dirac cone is independent from pseudopotentials. Velocity of Dirac fermions is nearly isotropic and equal to $7.6 \times 10^5 m/s$ comparable with that in graphene sheet. Electron effective mass is equal to $0.76 m_e$ in $\Gamma \to K$ direction and $0.56 m_e$ in $\Gamma \to M$ direction. For hole effective mass, it is equal to $0.33 m_e$ in $M \to K $ direction and $0.96 m_e$ in $M \to \Gamma $ direction. Analyzing of electron density of states, DOS, reveals that the valance and conduction band edges are composed of hybridization of p-orbitals of carbon and nitrogen atoms. There is an anisotropy in contribution of atoms in the VBM and CBM so that the VBM is composed of equal contribution of nitrogen and carbon p-orbitals. On contrary, the CBM is belonging to p-orbitals of carbon atoms and there is no contribution from nitrogen ones. Dirac point is also appeared due to hybridization of p-orbitals of carbon and nitrogen atoms but with different weights. Bloch states plotted in Fig.\ref{band_pure}c,d shows that the VBM is coming from $\pi$-hybridization of carbon and nitrogen orbitals, whereas electronic wave function is distributed just on carbon atoms in the CBM as it was predicted by DOS

\par Dynamical stability of the structures can be evaluated using phonon band dispersion. Fig. S1 in supplementary shows phonon band structure of $C_3N$ monolayer. The sheet is dynamically stable because there is no imaginary phonon mode in the band structure. Results prepared using classical approach based on Tersoff parameters is also plotted to compare with DFT-based results. There is a flexural mode with quadratic dispersion around the $\Gamma$ point and two acoustic modes with linear dispersion around $\Gamma$ point. Some active Raman modes are observed in energies higher than $45 meV$. Phonon modes cross each other in $K$ point which is a property of hexagonal lattice. Results obtained from two different methods is the same in energies less than $100 meV$ and some deviation is observed in higher energies. However, qualitative behavior is the same and as a consequence, we have used classical method to compute phonon transmission spectrum in following.

\begin{figure}[ht] \label{Fig_Thermal0}
\includegraphics[height=90mm,width=85mm,angle=0]{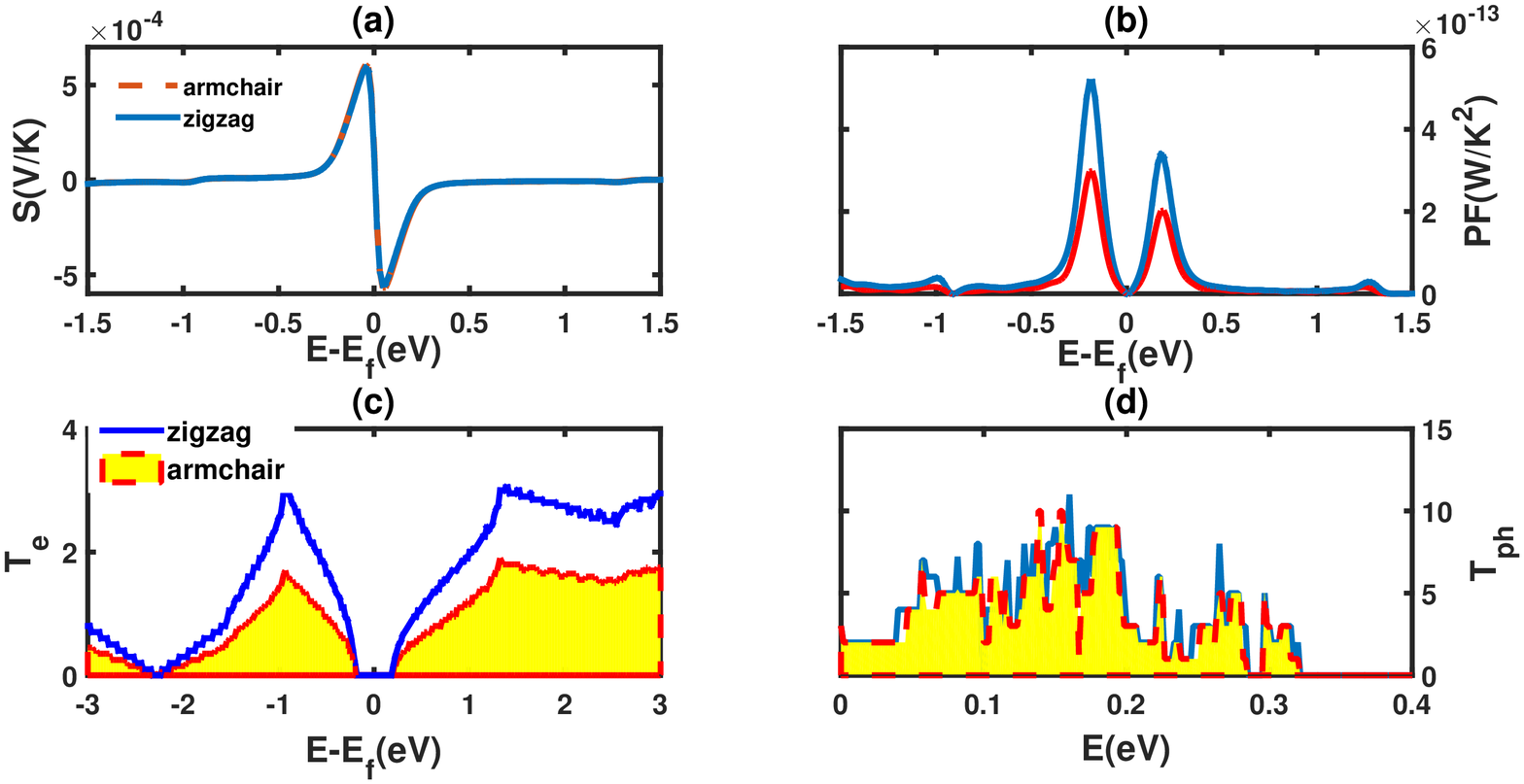}  
\caption{\label{Fig_Thermal0}  (a) Thermopower, and (b) power factor versus chemical potential. (c) and (d) electronic  and phonon transmission coefficient.}
\end{figure}

\begin{figure}[hb] \label{Fig_ZT0}
\includegraphics[height=85mm,width=85mm,angle=0]{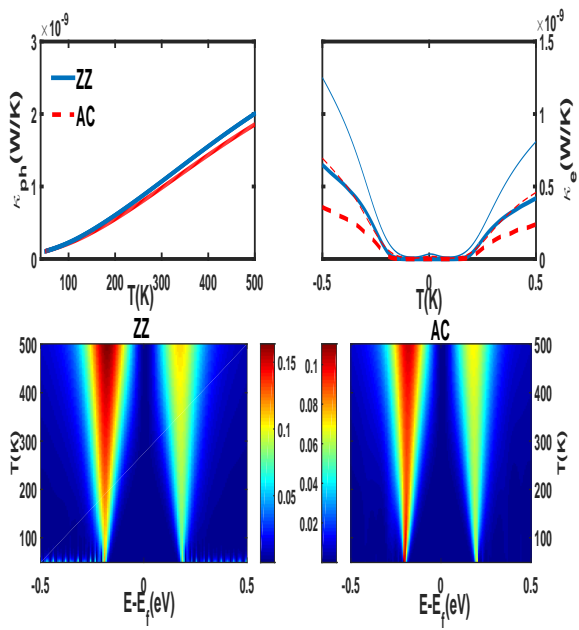}  
\caption{\label{Fig_ZT0}  (a) and (b) phonon and electron thermal conductance as a function of temperature. (c) and (d) shows color map of figure of merit as a function of temperature and chemical potential for strain along zigzag and armchair direction, respectively.}
\end{figure}

\par Two-dimensional materials are grown on a substrate. Difference between lattice parameter of the sheet and the substrate can induce some
compressive or tensile strain in the sheet resulting in changing electronic properties of it. For example, applying tensile strain can change
antimonene from an indirect semiconductor to direct one. Here, we investigate electronic and thermal properties of $C_3N$ sheet under compressive
and tensile uniaxial strain in the range of $-10\%<\epsilon<+10\%$.
Previous reports showed that the sheet is stable in the considered range in this research \cite{Zhou1, new1}.
uniaxial strain is applied along x-direction, zigzag direction (ZZ), and y-direction, armchair direction (AC),
as shown in Fig.\ref{gap_strain}a. For this purpose, we consider a supercell composed of 16 atoms with
orthorhombic shape. Variation of band gap versus strain is plotted in Fig.\ref{gap_strain}b. Electronic
response of $C_3N$ sheet to kind and direction of strain is very interesting and strange. First,
we analyze its response to strain in zigzag direction. Results obtained from both GGA and HSE approximations show
that the change of $E_g$ is significantly dependent on the kind of strain so that variation of $E_g$ is more intensive for compressive strain. Results from GGA approximation show that for compressive strain higher than $\geq0.8{\%}$,
 we observe a phase transition from semiconductor to metal. However, the phase transition is not supported by
 HSE calculations. Decrease of $E_g$ is slower for tensile strain. We fitted the data with a linear function with standard
 error less than $3{\%}$ and found that the slope of variation is three times higher in tensile strain. Band structure of
 the sheet under $\epsilon=-10{\%}$ and $+10{\%}$ is plotted in Fig.\ref{band_direction_strain}. A phase transition is observed from indirect semiconductor
 to   direct one upon applying tensile strain. This shows that $C_3N$ has potential for optoelectronic applications and LED
 devices using tensile strain. In addition, the range of $E_g$ obtained from HSE calculations shows that the $C_3N$ sheet can be
 used in solar cell devices due to closeness of obtained $E_g$ with silicone one. HSE-based data reveals that the band gap decreases $44{\%}$ for
  $\epsilon=-10{\%}$, while the reduction is about $18{\%}$ for $\epsilon=+10{\%}$. The anisotropy can be attributed to inconsistent change of lattice parameter perpendicular to strain direction.  For $\epsilon=10{\%}$, a reduction of $0.1\AA$ is measured in the lattice parameter, whereas we observe an increase of $0.2 \AA$ in lattice parameter for $\epsilon=-10{\%}$. The more intense of lattice parameter for compressive strain leads to more changes in bond length between atoms and as a consequence, more intense variation of $E_g$ as a function of compressive strain.
Change of bond length without and with applying strain is expressed in table. S1.

\par Response of the sheet to strain along armchair direction is different. We find that compressive strain induces a transition from indirect semiconductor to direct one. In addition, a transition from semiconductor to metal is observed for compressive strain higher than $4\%$. Results obtained from HSE calculations do not support the transition from semiconductor to metal, but change of band gap from indirect to direct is supported. In addition, variation of band gap versus compressive strain is more intensive for armchair direction. On contrary, variation of band gap for tensile strain is smother for armchair direction than zigzag one. Unlike zigzag direction, no transition from indirect to direct semiconductor is observed for tensile strain. HSE calculations support the observation. Band structure of $C_3N$ sheet under uniaxial strain along armchair direction is plotted in Fig.\ref{band_direction_strain}.

\subsection{Thermal properties}

\begin{figure}[ht] \label{PF_strain_ZZ}
\includegraphics[height=85mm,width=85mm,angle=0]{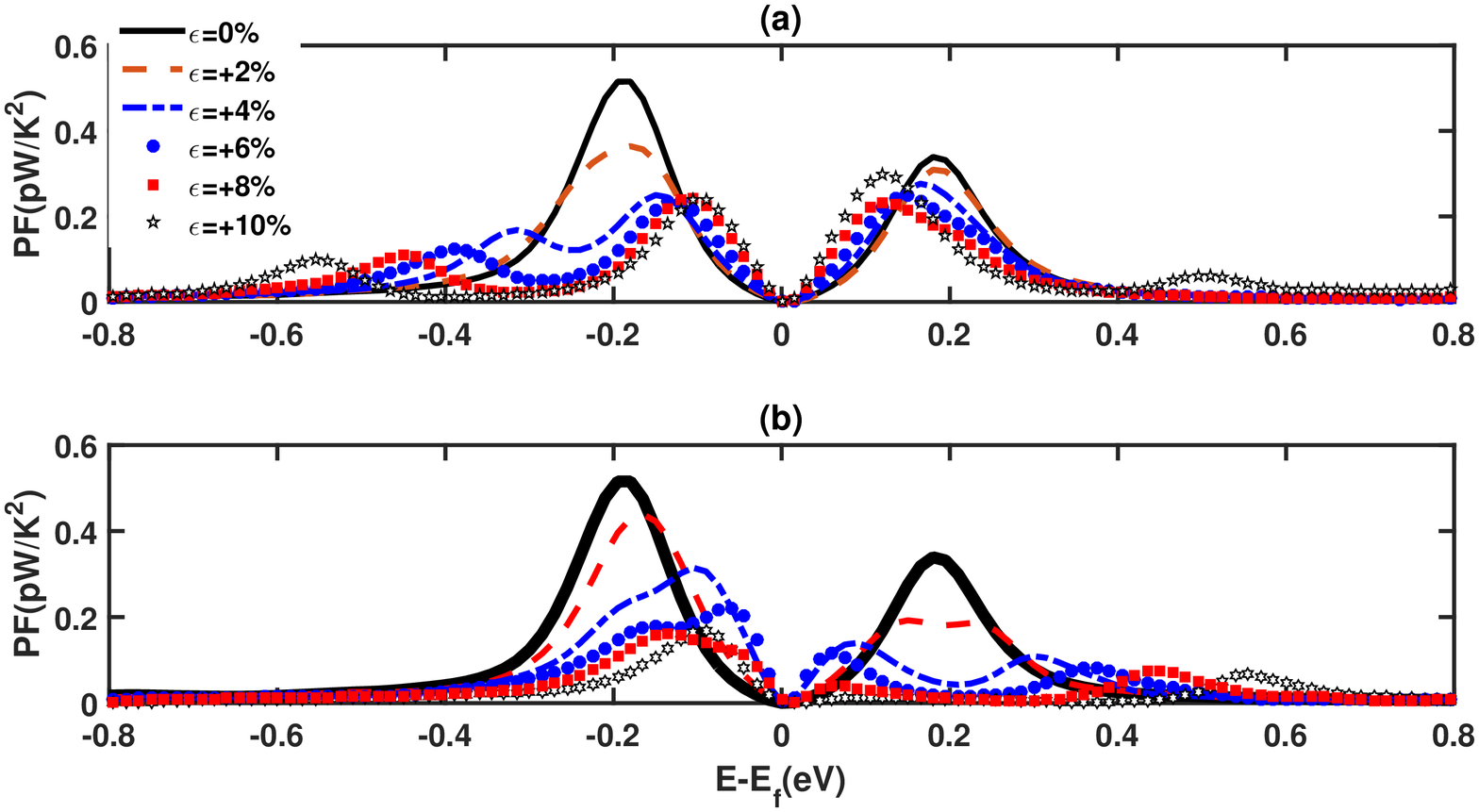}  
\caption{\label{PF_strain_ZZ} Power factor versus chemical potential for (a) tensile  and (b) compressive  strain along zigzag direction.}
\end{figure}

\par Thermopower of pristine $C_3N$ sheet is plotted in Fig.\ref{Fig_Thermal0}a. Thermopower is completely isotropic so that its behavior is independent from electron transport direction. A slight difference in its maximum value in two side of charge neutrally point is observed so that $S_{max}$ is slightly higher in p-doping ($\mu<0$) than n-doping ($\mu>0$). Maximum of thermopower is about $0.5 mV/K$ which is slightly lower than values reported for arsenene and antimonene \cite{Se1,Se2} and higher than $MoS_2$ \cite{Se3}.Unlike thermopower, power factor ($S^2G$) is an anisotropic function and dependent on the electron transport direction. It is clear from Fig.\ref{Fig_Thermal0}b that its value is higher in ZZ than AC. Maximum value of $PF$ obtained  for $C_3N$ is equal to values reported for phosphorene ($6 \times 10^{-13} W/K^2$) and lower than arsenene \cite{Se1}. Another difference between $PF$ of $C_3N$ and other 2D materials is existence of maximum of $PF$ close to the Fermi level attributed to the low band gap of the sheet. Peaks of $PF$ are correspond to band edges. Higher value of $PF$ in ZZ direction is directly related to higher value of electron transmission coefficient in that direction. $T_{el}$ in ZZ direction is twice of that in AC direction as it is shown in Fig.\ref{Fig_Thermal0}c. One reason can be related to the higher width of the unit cell used in ZZ direction. However, the main reason of the difference comes from the arrangement of the atoms along the transport direction. Phonon transmission spectrum is also direction dependent so in most range of energy, its value is higher in ZZ direction than AC one.  Maximum value of $T_{ph}$ calculated in this research is three times more than value reported for phosphorene, arsenene and $SnS$ monolayer \cite{Se1}. So, it is concluded that $C_3N$ is a very good thermal conductor like its counterpart graphene.

\begin{figure}[hb] \label{K_Z_strain}
\includegraphics[height=75mm,width=75mm,angle=0]{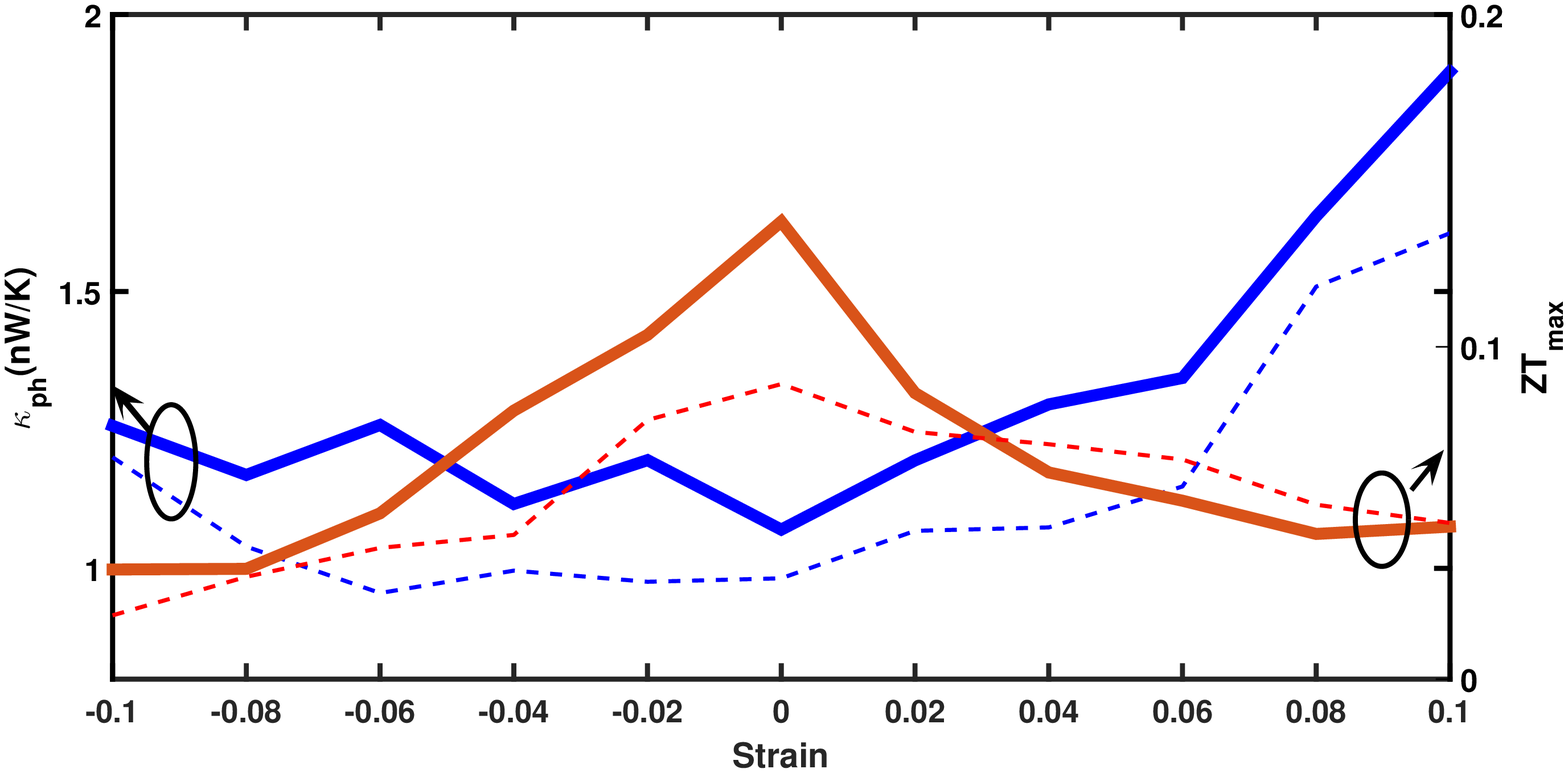}  
\caption{\label{K_Z_strain} Phonon thermal conductance and maximum of figure of merit for strain along zigzag (thick lines) and armchair (thin lines) directions.}
\end{figure}

\begin{figure}[ht] \label{ZT_ZZ_AC}
\includegraphics[height=80mm,width=90mm,angle=0]{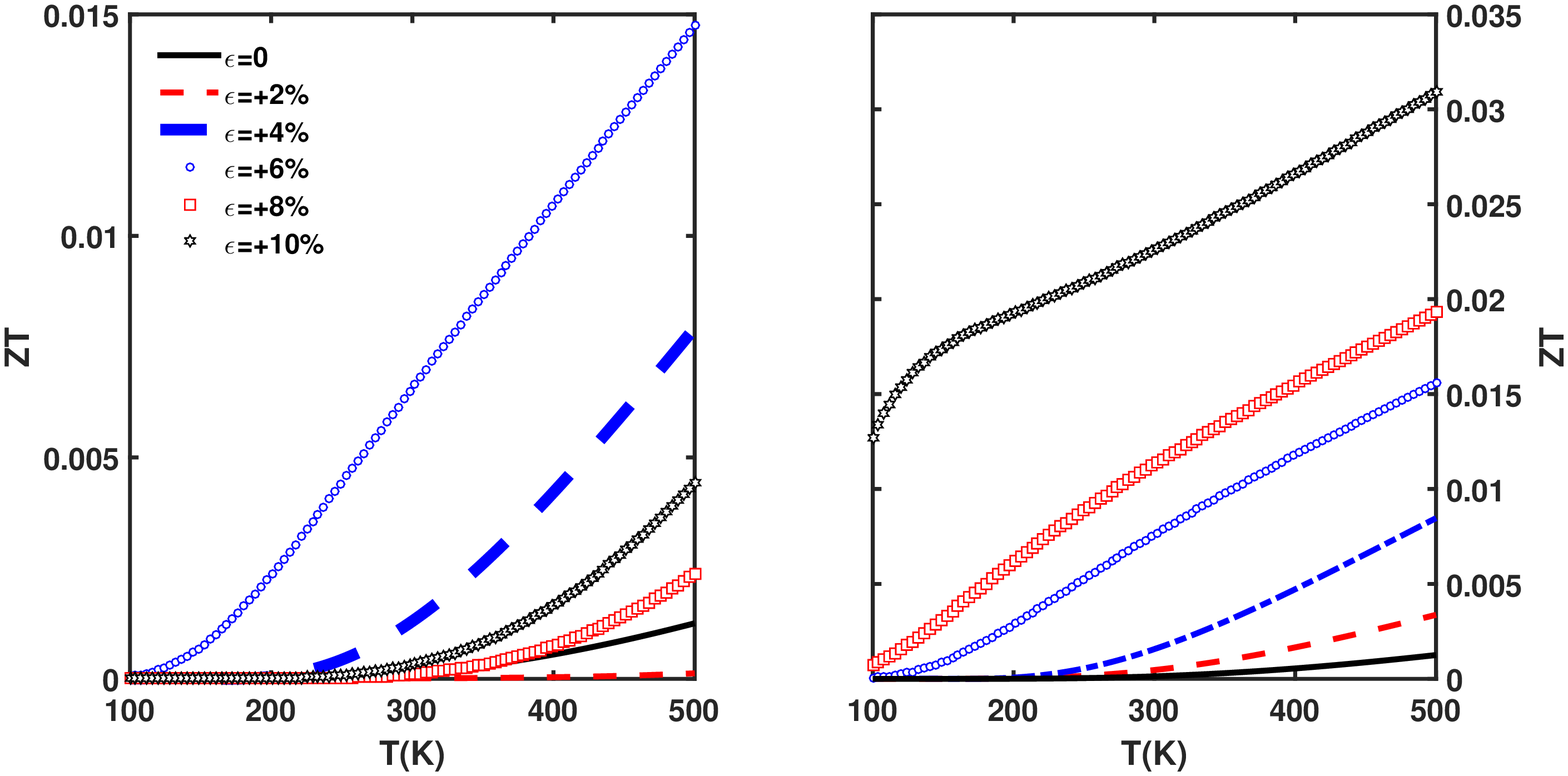}  
\caption{\label{ZT_ZZ_AC} Figure of merit versus temperature when strain is applied along zigzag direction while transport is along armchair direction. (a) Tensile  and (b) compressive strain.}
\end{figure}

\par Fig.\ref{Fig_ZT0} shows phonon and electron thermal conductance and figure of merit as a function of chemical potential correspond to n- or p-type doping. In low temperatures, $T<200 K$, phonon thermal conductance is isotropic and increase of temperature induces some anisotropy in it. $\kappa_{ph}$ is more in ZZ direction in high temperatures. Same results were reported in Refs.\cite{new1, new3} obtained from molecular dynamic simulations. Linear increase of $\kappa_{ph}$ in high temperatures is attributed to high energy phonon modes. Indeed, in high temperatures optical phonon modes are contribute in thermal transport. Because Debye temperature of $C_3N$ is much higher than arsenene, or antimonene, saturation of $\kappa_{ph}$ is not observed in $C_3N$ monolayer unlike arsenene or antimonene \cite{Se1, Se4}. However, thermal conductivity of the sheet decreases with increase of temperature \cite{new2, new3} Electron thermal conductance exhibits more intense anisotropy than phonon one. Like $\kappa_{ph}$, $\kappa_{e}$ calculated in ZZ direction is higher than AC one. Increase of temperature increases $\kappa_e$ which was predictable. $\kappa_e$  in AC direction at $400 K$ is equal to that for ZZ direction at $200 K$. Figure of merit shows tow main peaks one in p-type and the other in n-type doping. VBM and CBM has nearly equal distance from the Fermi level in GGA calculation so the position of ZT peaks is symmetric relative to the Fermi level. Increase of temperature widens the ZT peaks. It is observed that p-type doping can produce higher thermoelectric efficiency than n-type one. In addition, ZT is higher in ZZ direction so that $\frac{ZT_{max}(ZZ)}{ZT_{max}(AC)}=1.49$. Maximum of $ZT$ obtained for pristine $C_3N$ monolayer is lower than phosphorene, arsenene, or antimonene because of its higher thermal conductance. However, it is higher than the value was reported for graphene based on molecular dynamic simulations ($ZT=0.08$) \cite{ZT1}, or graphyne with $ZT=0.15$ based on first-principles calculations\cite{ZT1}.

\begin{figure}[ht] \label{ZT_AC_ZZ}
\includegraphics[height=70mm,width=85mm,angle=0]{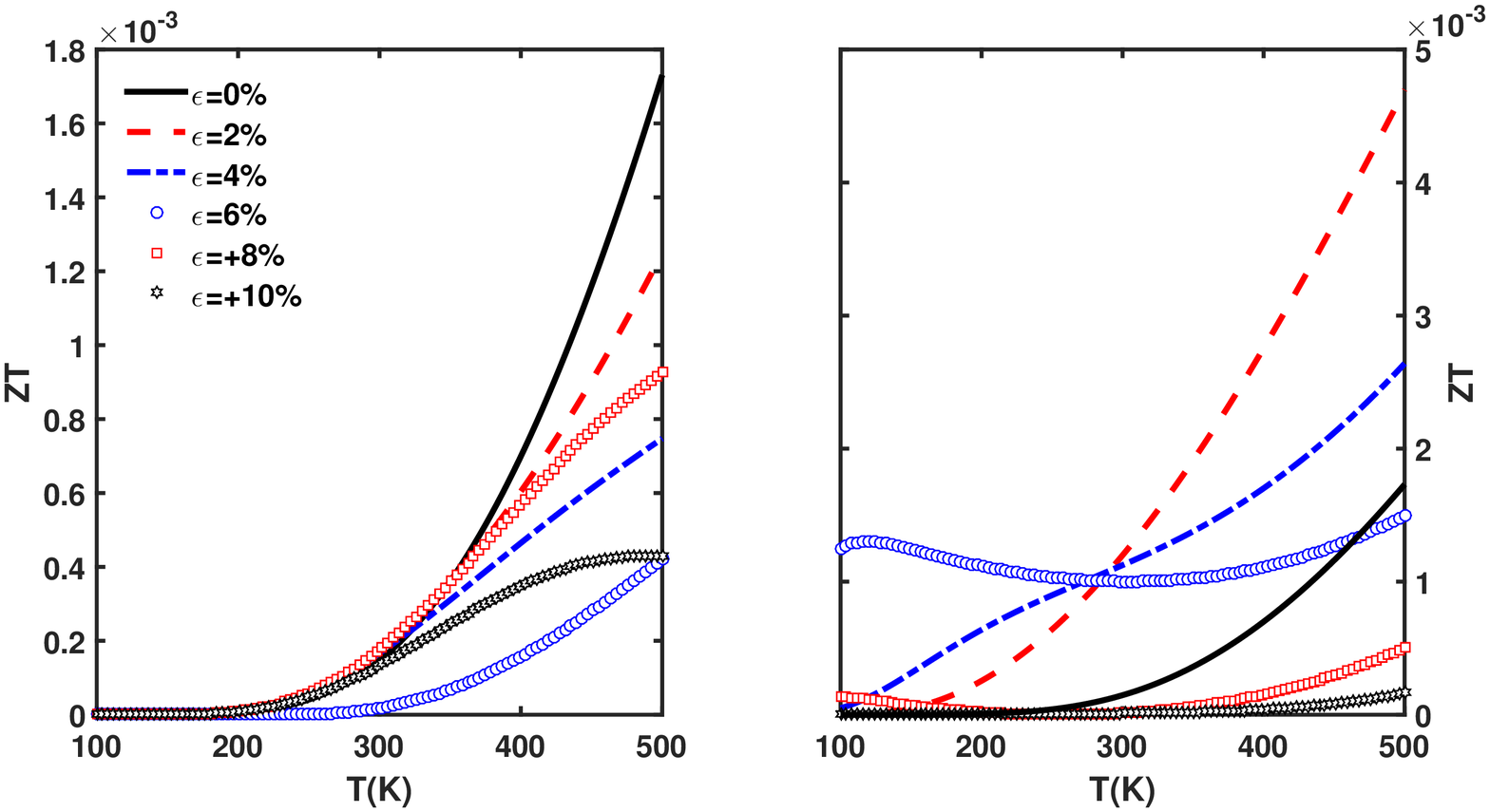}  
\caption{\label{ZT_AC_ZZ} Figure of merit versus temperature when strain is applied along armchair direction while transport is along zigzag direction. (a) Tensile  and (b) compressive strain.}
\end{figure}

\par In following, we analyze thermoelectric response of $C_3N$ sheet to uniaxial strain in different directions. Fig.\ref{PF_strain_ZZ} shows $PF$ of the sheet versus strain along ZZ direction. Increase of strain gives rise to reduction of $PF$. The reduction is more intensive in $n$ doping. Another point is appearance of side peaks by applying tensile strain in $p$ doping.  Reduction of band gap by strain reduces the gap in conductance and increases oscillation of thermopower and these effects induce some extra peaks in $PF$ spectrum.  About compressive strain, the reduction of $PF$ is more intensive in $n$ doping and side peaks are observed in $n$ doping unlike tensile case. Variations of $PF$ as a fuction of strain along AC direction is shows in Fig.{S4}. Unlike ZZ direction, tensile strain along AC direction has no significant effect on the magnitude of $PF$. The reason is attributed to the smooth change of the band gap versus tensile strain in AC as shown in Fig.\ref{gap_strain}b. For compressive strain higher than $4{\%}$, due to transition from semiconductor to metal, the curvature of $PF$ is significantly changed. The changes come from oscillations of the thermopower and gapless electrical conductance which are fundamental features of the metals. As an overall, we find that the $PF$ is reduced by applying strain.

\par Dependence of lattice thermal conductance and maximum of thermoelectric efficiency on the strain is plotted in Fig.\ref{K_Z_strain}. We observe a uniform increase of phonon thermal conductance with increase of tensile strain. Increase is more obvious in ZZ direction than AC one. On contrary, $\kappa_{ph}$ shows oscillatory behavior versus compressive strain along ZZ direction. For Ac strain less than $-6{\%}$, $\kappa_{ph}$ is almost constant and after that a linear increase is observed. Results show that the lattice thermal conductance is always higher in ZZ direction. Applying strain results in decrease of maximum of $ZT$ and this result is independent from strain direction. However, the reduction of $ZT$ is more in ZZ direction. Results show that maximum of figure of merit obtained from tensile strain along AC direction exceeds from that of ZZ direction for $\epsilon>+3{\%}$. $ZT_{max}$ becomes independent from strain direction in high strain. Based on the results, it is clear that strain engineering can be used as valuable tool to tune thermal and thermopower properties of the $C_3N$ sheet. We believe that thermoelectric efficiency of $C_3N$ sheet can be more than values reported in this work if one uses quantum mechanical approach to compute phonon transmission coefficient. Here, we compute phonon transmission coefficient in $\Gamma$ point and averaging of it on the k-space can significantly reduce $T_{ph}$ and as a result increase $ZT$ of the sheet.

\par Until now, we study transport properties of $C_3N$ sheet under condition that strain and transport are parallel. In following, we consider situations in which transport direction and strain are perpendicular to each other i.e. the strain is along $x$ direction (ZZ) while the transport is along $y$ direction (AC) and vice versa. Fig.\ref{ZT_ZZ_AC} shows $ZT$ versus temperature for strain applied along ZZ direction and transport along AC one. We find that the $ZT$ shows nonlinear behavior for tensile strain so that it reduces for $\epsilon=+2{\%}$ and then significantly increases for $3{\%}<\epsilon<6{\%}$ . Threshold temperature in which $ZT$ is higher than $0.001$ is lowered by increase of strain. In addition, we observe a linear behavior for $ZT$ in high temperatures. Although more increase of strain reduces the magnitude of $ZT$, however, it is still higher than $\epsilon=0$ case. Upon applying compressive strain, a substantial increase in $ZT$ is observed that unlike tensile case it never decreases with increase of strain magnitude. Our investigations reveal that for $\epsilon=-10{\%}$ and at $500K$, $ZT$ is 23 times higher than an unstretched sheet in the same temperature.

\par Fig.\ref{ZT_AC_ZZ} shows thermoelectric efficiency of the sheet under conditions that the strain as applied along AC direction, while transport is happened along ZZ direction. Unlike previous case, $ZT$ of unstretched sheet is more in room temperature than that under tensile strain. In addition, behavior of $ZT$ is nonlinear at high temperatures. On the other hand, compressive strain results in the increase of $ZT$. Finite values of $ZT$ at low temperatures and under high tensile strain is a result of transition from semiconductor to metal because of gapless electrical conductance. Obtained results show that magnitude of $ZT$ is lower in this case than the strain along ZZ direction and transport along AC direction. Results prove that thermoelectric abilities of the $C_3N$ sheet are strongly dependent on the strain engineering and transport direction and this phenomenon makes it a promising candidate for thermoelectric applications.

\section{\label{sec:level4}Conclusions}
Ponyaniline sheet is a recently synthesized two-dimensional monolayer with a hexagonal unit cell composed of six carbon and two nitrogen atoms with empirical formula of $C_3N$.
We have investigated the electronic and thermal properties of $C_3N$ sheet under external strain using density functional theory combined with Green function formalism in the linear response regime. We find that the band gap of the sheet decreases with increase of strain magnitude and the slope of the reduction is dependent on the strain direction and kind (tensile or compressive). Our investigations show that thermopower is independent from strain direction, along zigzag or armchair direction, but power factor and as a result figure of merit are strongly dependent. External strain can reduce the thermoelectric efficiency of the sheet when the strain and electron transport are parallel. We show that thermoelectric efficiency of the sheet can be noticeably modified with control of strain and transport direction. Results show that the thermoelectric efficiency increases significantly at room temperature under conditions that strain and transport directions are perpendicular to each other.

\section{Acknowledgment}
We are thankful to the Research Council of the University of Guilan for the partial support of this research.

\end{document}